\documentclass[prd,twocolumn,nopacs,floatfix,amsmath,nofootinbib,amssymb,floatfix]{revtex4}
\usepackage{graphicx,color,dcolumn,booktabs,bm}
\usepackage{longtable,lscape}
\usepackage{pdfpages}
\usepackage{txfonts}
\usepackage{overpic}
\usepackage{amssymb}
\usepackage{indentfirst}
\usepackage{feynmf}   
\usepackage{slashed}  
\usepackage{cases}
\usepackage{color}
\usepackage{multirow}
\usepackage{threeparttable}
\usepackage{epstopdf}
\usepackage{enumerate}
\usepackage{amsmath}
\usepackage{graphicx,color,dcolumn,booktabs,bm}
\usepackage[colorlinks, citecolor=blue,anchorcolor=red,menucolor=red, linkcolor=red,filecolor=red,urlcolor=blue,frenchlinks=red]{hyperref}

\graphicspath{{Figures/}} %

\begin{document}

\title{Spectral Analysis and Decay Mechanisms of $1^{-+}$ Hybrid States in Light Meson Sector}

\author{Fu-Yuan Zhang$^{1}$}\email{fyzhang1110@163.com}
\author{Qi Huang$^{2}$}\email{06289@njnu.edu.cn}
\author{Li-Ming Wang$^{1}$\footnote{Corresponding author}}\email{lmwang@ysu.edu.cn}
\affiliation{ $^1$Key Laboratory for Microstructural Material Physics of Hebei Province, School of Science, Yanshan University, Qinhuangdao 066004, China\\
$^2$Department of Physics, Nanjing Normal University, Nanjing 210023, People’s Republic of China
}

\date{\today}

\begin{abstract}
The exploration of exotic mesons, which transcend the conventional quark-antiquark framework, is pivotal for advancing our understanding of QCD and the strong interaction. Among these, states possessing the quantum numbers $J^{PC}=1^{-+}$, such as $\pi_1(1600)$, $\pi_1(2015)$, and the recently discovered $\eta_1(1855)$, have attracted significant attention due to their potential hybrid nature, which involves gluonic excitations. However, the physical interpretation of these states remains contentious, primarily due to inconsistencies between theoretical predictions and experimental observations regarding their decay widths and mass spectra. To address these challenges, we employ a potential model inspired by SU$(3)$ lattice gauge theory to calculate the masses of light-flavor $1^{-+}$ hybrid states and utilize the constituent gluon model to analyze their strong decay properties. Our mass calculations suggest that while the $\pi_1(1600)$ and $\eta_1(1855)$ masses are in close agreement with the predicted hybrid states, the corresponding decay widths for $\pi_1(1600)$ and $\pi_1(2015)$ do not support their classification as hybrids, implying alternative interpretations, such as tetraquark or molecular configurations. Conversely, the $\eta_1(1855)$ is consistent with a mixed hybrid state composed of $(u\bar{u}+d\bar{d})g/\sqrt{2}$ and $s\bar{s}g$ components, with mixing angles ranging from $17.7^\circ$ to $84.2^\circ$. Additionally, we predict the masses and decay widths of yet-unobserved isoscalar and excited hybrid states, providing valuable targets for future experimental searches. This study not only clarifies the hybrid nature of certain exotic mesons but also contributes to the systematic classification of hybrid states within the meson nonet, thereby enhancing our comprehension of exotic hadronic states.
\end{abstract}



\maketitle

\section{Introduction}\label{sec1}

Approximately five decades ago, physicists formulated Quantum Chromodynamics (QCD) to describe the strong interaction \cite{tHooft:1972tcz,Fritzsch:1973pi,Gross:1973id,Politzer:1973fx,Gross:2022hyw}. This development has significantly advanced our understanding of most hadrons and established the framework for hadron spectroscopy. However, our understanding of the gluon field's role in the low-energy regime of QCD remains limited. Consequently, the study of conventional hadrons, where mesons consist of a quark and an antiquark, and baryons are composed of three quarks, appears insufficient. In response, the investigation of hybrid states, where gluons are incorporated as dynamic degrees of freedom, is crucial. This research is guided by two fundamental principles of QCD namely, asymptotic freedom \cite{Gross:1973id,Politzer:1973fx} and local color conservation \cite{Wilson:1974sk} and is essential for a deeper understanding of non-perturbative QCD phenomena \cite{Fritzsch:1973pi,Freund:1975pn}.

The exploration of exotic mesons, which go beyond the conventional quark-antiquark framework, is crucial for advancing our understanding of QCD and the strong interaction. These mesons, characterized by unconventional quantum numbers, challenge our current theories and provide a unique opportunity to study the role of gluonic excitations. Among them, states with $J^{PC}=1^{-+}$, such as $\pi_1(1600)$ \cite{E852:1998mbq}, $\pi_1(2015)$ \cite{E852:2004gpn} and the recently observed $\eta_1(1855)$ \cite{BESIII:2022riz,BESIII:2022iwi}, have drawn significant attention due to their potential hybrid nature. However, interpreting the properties of these states remains a contentious issue, largely due to discrepancies between theoretical predictions and experimental observations.

Several collaborations \cite{IHEP-Brussels-LosAlamos-AnnecyLAPP:1988iqi,Aoyagi:1993kn,E852:1997gvf,VES:2001rwn,CrystalBarrel:1998cfz} have reported the observation of $\pi_1(1400)$, but its existence as a resonance remains controversial \cite{Meyer:2010ku}. The $\pi_1(1600)$ has also been observed in various decay channels, including $\eta^\prime\pi$ \cite{Baker:2003jh}, $\rho\pi$ \cite{Khokhlov:2000tk}, $f_1(1285)\pi$ \cite{COMPASS:2009xrl}, and $b_1(1235)\pi$ \cite{CLEO:2011upl}. Over the years, many theorists have interpreted $\pi_1(1600)$ as a hybrid \cite{Narison:1999hg,Shastry:2022mhk,Chen:2023ukh,Iddir:1988jd,Benhamida:2019nfx,Eshraim:2020ucw}. Additionally, $\pi_1(1600)$ has been considered a candidate for a tetraquark state \cite{Chen:2008qw}, given that gluons in a hybrid state can easily split into a quark-antiquark pair ($q\bar{q}$). Furthermore, some theories suggest that $\pi_1(1600)$ may be a mixing state of a molecular state and a tetraquark or hybrid \cite{Narison:2009vj}. The $\pi_1(2015)$, observed exclusively in $b_1(1235)\pi$ \cite{E852:2004rfa} and $f_1(1285)\pi$ \cite{E852:2004gpn} by the BNL E852 experiment, remains poorly understood due to a lack of comprehensive experimental data. Lattice QCD calculations of radial excited states identify $\pi_1(2015)$ as a promising candidate for the first excited state of a hybrid \cite{Dudek:2010wm}.

The $\eta_1(1855)$, discovered through partial-wave analysis of $J/\psi \to \gamma \eta_1(1855) \to \gamma \eta \eta^\prime$ \cite{BESIII:2022riz,BESIII:2022iwi}, is the first isoscalar particle with quantum numbers $J^{PC} = 1^{-+}$. This discovery offers a unique opportunity to study exotic states. Extensive theoretical investigations have been carried out on the $\eta_1(1855)$. Within the framework of QCD sum rules, it has been interpreted as a four-quark state \cite{Wan:2022xkx}. Additionally, the $\eta_1(1855)$ has been studied as a molecular state \cite{Dong:2022cuw,Yang:2022rck,Yan:2023vbh} and as a hybrid state candidate using flux tube models \cite{Qiu:2022ktc}, QCD sum rules \cite{Chen:2022qpd}, and effective Lagrangian techniques \cite{Shastry:2022mhk}. Like $\pi_1(1600)$, the physical nature of $\eta_1(1855)$ remains unresolved.

The limited number of discovered exotic particles and the scarcity of available data present significant challenges in accurately classifying these states. To address these challenges, we adopt a potential model inspired by SU(3) lattice gauge theory, which has demonstrated success in capturing gluonic contributions to hadron interactions, to calculate the masses of light-flavor $1^{-+}$ hybrid mesons. Additionally, the constituent gluon model is employed to analyze their strong decay properties, offering a framework to investigate the interplay of gluonic and quark-antiquark dynamics in hybrid mesons. We also predict the masses and decay widths of yet undiscovered isoscalar and excited hybrid states, which provide valuable targets for experimental exploration. This study not only clarifies the hybrid nature of specific exotic mesons but also enhances the systematic classification of hybrid states within the meson nonet. Furthermore, conventional mesons are known to follow Regge trajectories \cite{Anisovich:2000kxa}. Investigating whether hybrid states also adhere to this pattern is crucial for establishing a complete SU(3) multiplet, as discussed by other authors \cite{Chen:2023ukh,Yan:2023vbh,Qiu:2022ktc}. By addressing these unresolved issues, our work contributes to a deeper understanding of the internal structure of exotic hadronic states and the dynamics of the strong interaction.

This paper is organized as follows: In Section \ref{sec2}, we introduce the model for calculating the masses of $1^{-+}$ hybrid states and present theoretical masses for various flavor compositions. Section \ref{sec3} describes the constituent gluon model used to calculate the decay widths of hybrid states. In Section \ref{sec4}, we discuss the decay widths of both discovered and undiscovered $1^{-+}$ particles, along with their dominant decay channels. Finally, Section \ref{sec5} concludes with a discussion and summary.

\section{Masses of $1^{-+}$ hybrid }\label{sec2}

The quark potential model has proven highly effective in accurately predicting the masses of conventional mesons and baryons, prompting its application to hybrid systems. Recent advancements in lattice QCD have enabled the modeling of effective potentials for excited gluon fields, which facilitate interactions between quarks \cite{UKQCD:1998zbe,Karbstein:2018mzo,Capitani:2018rox,Schlosser:2021wnr}. These developments provide a foundation for the computation of hybrid meson masses within the  $1^{-+}$ sector \cite{Chen:2023ukh}. The effective potential for the lowest hybrid state with $J^{PC}=1^{-+}$ in the light-flavor sector is modeled as follows:
\begin{equation}
V_{q\bar{q}g}(r) = \frac{A_1}{r} + A_2 r^2 + V_0 + \xi_2 \sqrt{\frac{b}{\xi_1}},
\label{eq1}
\end{equation}
where $q$ and $\bar{q}$ denote the light ($u$, $d$) and strange ($s$) quarks. The parameters are set as $A_1=0.0958$ GeV, $A_2=0.01035$ GeV$^3$, $b=0.165$ GeV$^2$, $\xi_1=0.04749$, and $\xi_2=0.5385$. The constant $V_0$ varies with flavor: $V_0^{(q\bar{q})}=-0.48$ GeV, $V_0^{(q\bar{s})}=-0.40$ GeV, and $V_0^{(s\bar{s})}=-0.33$ GeV.

We solve the spinless Salpeter equation:
\begin{equation}
H = \sqrt{p_q^2 + m_q^2} + \sqrt{p_{\bar{q}}^2 + m_{\bar{q}}^2} + V_{q\bar{q}g}(r),
\label{eq2}
\end{equation}
where $p_q$ and $p_{\bar{q}}$ are the momenta of the quark and antiquark, respectively, and $m_q$, $m_{\bar{q}}$ their masses.

The wave functions are approximated using the Simple Harmonic Oscillator (SHO) basis:
\begin{equation}
\psi_{nl}(r,\beta) = \beta^{3/2} \sqrt{\frac{2(2n-1)!}{\Gamma(n+l+\frac{1}{2})}} (\beta r)^l e^{-\frac{\beta^2 r^2}{2}} L^{l+1/2}_{n-1}(\beta^2 r^2),
\label{eq03}
\end{equation}
where $L^{\alpha}_n$ are the associated Laguerre polynomials, and $\beta$ is the oscillator parameter specific to each flavor configuration: $\beta^H_{q\bar{q}}=0.264 ~\text{GeV}, \beta^H_{q\bar{s}}=0.271 ~\text{GeV}$ and $\beta^H_{s\bar{s}}=0.277 ~\text{GeV}$.

By solving the expectation value of energy, the corresponding mass ${M}_{nl}$ is obtained as 
\begin{equation}
M_{nl}(\beta)\equiv
E_{nl}(\beta) =\langle\psi_{nl}|H|\psi_{nl}\rangle. \label{eq3}
\end{equation}

\begin{table}[htbp]
\caption{ The masses of hybrid states with the radial quantum number $n$. All masses are in units of GeV. }\label{table01}
\renewcommand\arraystretch{1.2} 
\begin{tabular*}{86mm}{@{\extracolsep{\fill}}cccccc}
\toprule[1pt]\toprule[1pt]
$n$          & 1         & 2               & 3            & 4         & 5               \\
\cline{1-6}
$q\bar{q}g$  & 1.668     & 2.193           & 2.648        &3.061      & 3.444        \\
$q\bar{s}g$  & 1.851     & 2.355           & 2.799        & 3.204     & 3.582          \\
$s\bar{s}g$  & 2.022     & 2.507           & 2.94         & 3.337     & 3.709           \\
            \bottomrule[1pt]\bottomrule[1pt]
\end{tabular*}
\end{table}

In this study, we use a potential model derived from lattice QCD to calculate the masses of light-flavor $1^{-+}$ hybrid states. The effective potential includes contributions from Coulomb-like terms, confinement effects, and gluonic excitations. The spinless Salpeter equation is solved using a harmonic oscillator wavefunction basis, allowing us to obtain the energy eigenvalues. The computed masses are presented in Table \ref{table01} and show close agreement with experimental data for $\pi_1(1600)$ and $\eta_1(1855)$, which correspond to $q\bar{q}g$ and $q\bar{s}g$ configurations, respectively. However, deviations are observed for $\pi_1(2015)$, suggesting it may belong to a different excitation class.

\begin{figure}
    \centering
    \includegraphics[width=0.85\linewidth]{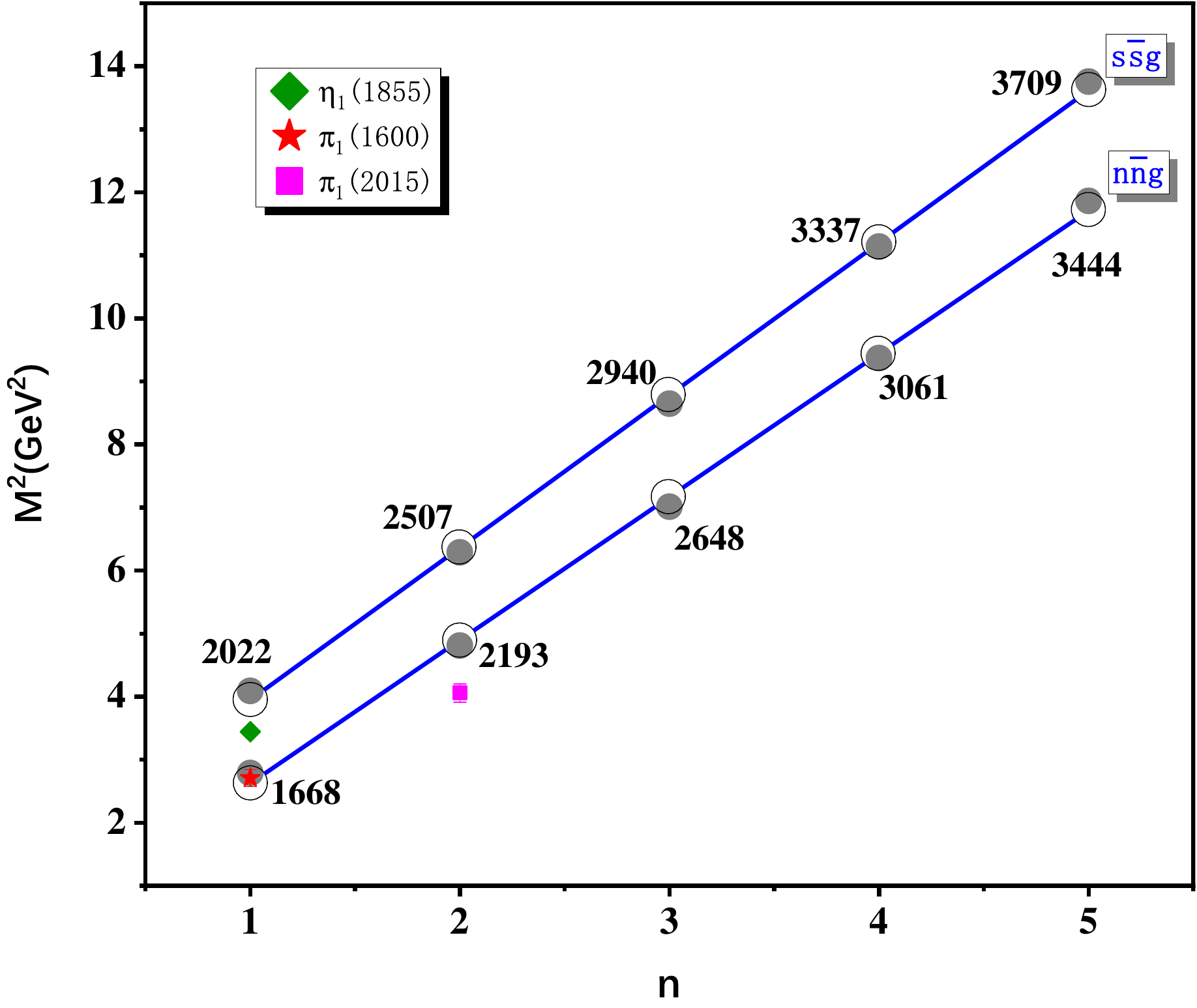}
    \caption{(color online). The analysis of the ($n,M^2$) trajectories for the $1^{-+}$ states. The trajectory slopes are 2.41 and 2.27 for the $s\bar{s}g$ and $n\bar{n}g$ states, respectively. Here,the solid circles with gray denote calculated values of phenomenological theory and circles denote the theoretical values of Reggie trajectory, while the error bar correspond to the experimental data listed in  Particle Data Grou (PDG) \cite{ParticleDataGroup:2024cfk}.
}
    \label{xianxing}
\end{figure}

Additionally, we analyze the mass spectrum using Regge trajectories (see Fig. \ref{xianxing}), which describe the linear relationship between squared masses of mesons and their quantum numbers. The observed adherence of hybrid mesons to Regge trajectories, similar to conventional mesons, suggests that hybrid states may exhibit regularities comparable to traditional meson families. This finding suggests that hybrids may exhibit regularities akin to those of ordinary mesons, paving the way for the systematic classification of hybrid mesons as part of a broader meson family.

The fitting result equation corresponding to Fig. \ref{xianxing} :
\begin{eqnarray}
M_{s\bar{s}g}^{2}=&M_{s0}^2+(n-1)\mu_1,            \nonumber\\
M_{q\bar{q}g}^{2}=&M_{q0}^2+(n-1)\mu_2, \label{eq4}
\end{eqnarray}
where $M_{s0}^2$ and $M_{q0}^2$ represent the ground state masses of $s\bar{s}g$ and $q\bar{q}g$, respectively.

The Reggie trajectory exhibited by mesons in  Ref. \cite{Anisovich:2000kxa} is
\begin{eqnarray}
M_{q\bar{q}}^{2}=M_0^2+(n-1)\mu, \label{eq5}
\end{eqnarray}
where $\mu_1=2.41$, $\mu_2=2.27$ and $\mu=1.25$.

In isospin partner mesons, the $s\bar{s}g$ and $n\bar{n}g$ flavor configurations often mix, with $n\bar{n}g$ denoting $(u\bar{u}+d\bar{d})/\sqrt{2}g$. The calculated mass of $s\bar{s}g$ is slightly higher than that of $\eta_1(1855)$, while the mass of $n\bar{n}g$ is slightly lower. This suggests that the $\eta_1(1855)$ state could be a hybrid of these two configurations. In Section \ref{sec4}, we analyze the strong decay behavior of the $\eta_1(1855)$, treating it as a mixture of $s\bar{s}g$ and $n\bar{n}g$.

\section{Decay model of the hybrid state}\label{sec3}

\begin{figure}
    \centering
    \includegraphics[width=0.7\linewidth]{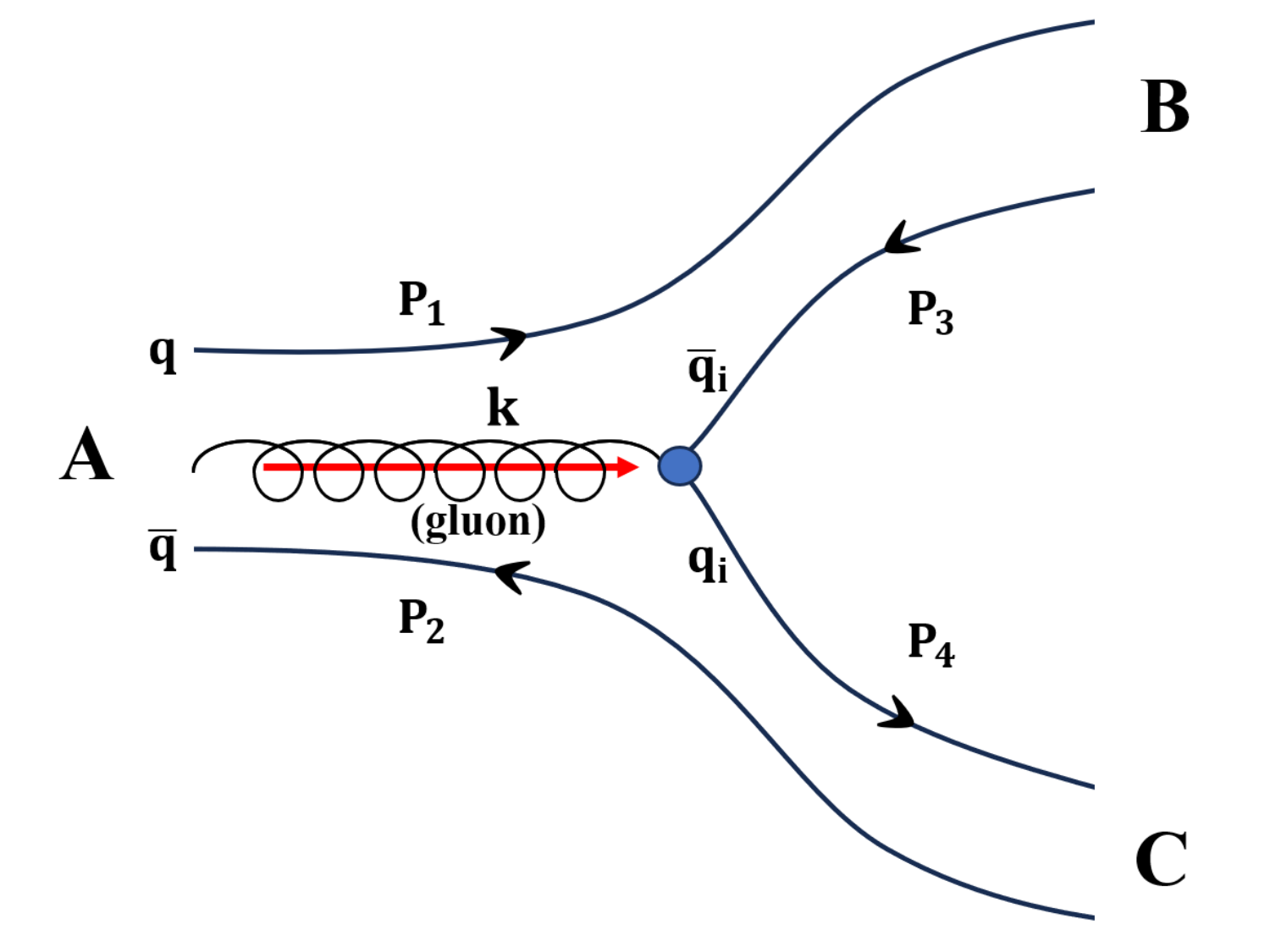}
    \caption{The mechanism of a hybrid state A decaying into B and C mesons through the constituent gluon model.}
    \label{model}
\end{figure}

The decay properties of hybrid states are essential for understanding their structure and classification.
Several theoretical models have been developed to investigate the decay widths of hybrid states, including the flux tube model \cite{Page:1998gz,Isgur:1985vy,Close:1994hc,Barnes:1995hc}, the constituent gluon model \cite{LeYaouanc:1984gh,Iddir:1988jd,Ishida:1991mx,Kalashnikova:1993xb,Swanson:1997wy,Iddir:2000yb,Ding:2006ya,Iddir:2007dq,Benhamida:2019nfx,Farina:2020slb,Tanimoto:1982eh,Tanimoto:1982wy}, QCD sum rules \cite{DeViron:1984svx,Zhu:1998sv,Zhang:2002id,Chen:2010ic,Huang:2010dc,Huang:2016upt}, and lattice QCD \cite{McNeile:2002az,McNeile:2006bz,Woss:2020ayi}. 
In this study, we employ the constituent gluon model, which describes hybrid mesons as quark-antiquark pairs coupled to a transverse electric (TE) gluon with quantum numbers $J_g^{PC} = 1^{+-}$ \cite{Chen:2023ukh}. This TE gluon functions as a dynamic degree of freedom, capable of annihilating into a new quark-antiquark pair. The resulting pair then combines with the original quarks to form two mesons, as depicted in Fig. \ref{model}. This mechanism provides a comprehensive description of the decay process of hybrid states into meson pairs.

For the representation of the hybrid states the following notations
are used:

L$_{\text{g }}$\emph{\ \ }: the relative orbital momentum
of the gluon in the \emph{q$\bar{q}$} center of mass;

$L_{q\bar{q}}$ \emph{\ }: the relative orbital momentum between
\emph{q} and \emph{$\bar{q}$};

$S_{q\bar{q}}$ \emph{\ }: the total quarks spin;

$J_{g}$ \emph{\ }: the total gluon angular momentum;

\begin{eqnarray}
J_{g}&\equiv&L_{\text{g }}+1 \nonumber,\\ \label{eq6}
L&\equiv&L_{q\bar{q}}+J_{g} =L_{q\bar{q}}+L_{\text{g }}+1.
\end{eqnarray}

Considering the gluon moving in the framework of the $q\overline{q}$
pair, the Parity of the hybrid will be \cite{Iddir:2007dq}: 
\begin{equation}
\label{eq07}
P=\left(-\right)^{L_{q\bar{q}}+L_{g}}.
\end{equation}
The charge conjugation of hybrid states is expressed as: 
\begin{equation}
\label{eq08}
C=\left(-\right)^{L_{q\bar{q}}+S_{q\bar{q}}+1}.
\end{equation}

$S_{q\bar{q}}$ can be either 0 or 1, Eq. \ref{eq07} and Eq. \ref{eq08} imply that both $L_{q\bar{q}}+L_{\text{g}}$ and $L_{q\bar{q}}+S_{q\bar{q}}$ must be odd for $J^{PC}=1^{-+}$. Under the constraints of the Clebsch-Gordan coefficients in Eq. \ref{eq12}, the lightest states of $J^{PC}=1^{-+}$ hybrid is presented in Table \ref{table02}, which is categorized into two general classes:
 $L_{q\bar{q}}=0$ and $L_{g}=1$ which is usually  referred as the gluon-excited hybrid
(GE hybrid), and $L_{q\bar{q}}=1$ and $L_{g}=0$ which is usually  referred as  the quark-excited hybrid (QE hybrid). 

Within the degree of freedom of the transverse electric (TE) gluon, the $1^{-+}$ hybrid state corresponds to the gluon-excited state (GE hybrid) \cite{LeYaouanc:1984gh}.

\begin{table}[htbp]
\caption{ The quantum numbers of $J^{PC}=1^{-+}$ $\,q \bar q g$ hybrid mesons}\label{table02}
\renewcommand\arraystretch{1.2} 
\begin{tabular*}{86mm}{@{\extracolsep{\fill}}cccccccc}
\toprule[1pt]\toprule[1pt]
$P$  & $C$ & $L_{q\bar{q}}$  & $L_{\text{g }}$            & $J_{g}$    & $S_{q\bar{q}}$  & $L$  & $J$ \\
$-$  & $+$ & 0               & 1                          & 0          & 1               & 0  & 1 \\
$-$  & $+$ & 0               & 1                          & 1          & 1               & 1 & 1 \\
$-$  & $+$ & 0               & 1                          & 2          & 1               & 2 & 1 \\
$-$  & $+$ & 1               & 0                          & 1          & 0               & 1 & 1 \\             \bottomrule[1pt]\bottomrule[1pt]
\end{tabular*}
\end{table}


 In QCD theory, the Hamiltonian describing the transition of a hybrid state into two conventional mesons is as follows:
\begin{equation}
{H}_I=g\int\textup{d}^3\vec{x}~\bar{\psi}(\vec{x})\gamma_\mu\frac{\lambda^a}{2}\psi(\vec{x})A_a^\mu(\vec{x}).  \label{eq9}
\end{equation}
The operator relevant to the decay can be expressed by operators that encompass the annihilation of gluons and the creation of quark-antiquark pairs:
\begin{eqnarray}
\label{eq10}
H_I=g\sum_{s,s^{\prime},\lambda,c,c^{\prime},c_{g}}\int\frac{d^3\mathbf{p}\;d^3\mathbf{p}^{\prime}d^3\mathbf{k}}{\sqrt{2\omega_g}(2\pi)^9}(2\pi)^3\delta^{3}
(\mathbf{p}-\mathbf{p^{\prime}}-\mathbf{k}) \nonumber \\
\times\bar{u}_{\mathbf{p}sc}\gamma_{\mu}\frac{\lambda^{c_g}_{c,c^{\prime}}}{2}v_{-\mathbf{p}^{\prime}s^{\prime}c^{\prime}}b^{\dagger}_{\mathbf{p}sc}
d^{\dagger}_{-\mathbf{p}^{\prime}s^{\prime}c^{\prime}}a^{c_g}_{\mathbf{k}\lambda}\varepsilon^{\mu}(\mathbf{k},\lambda).
\end{eqnarray}
 The flavor index of the quark, $c_g=1,2,\cdot\cdot\cdot8$, has been omitted and the creation and annihilation operators satisfy the relationship:
\begin{eqnarray}
\nonumber\{b_{\mathbf{p}sc},b^{\dagger}_{\mathbf{p}^{\prime}s^{\prime}c^{\prime}}\}&=&\{d_{\mathbf{p}sc},d^{\dagger}_{\mathbf{p}^{\prime}s^{\prime}c^{\prime}}\}=(2\pi)^3\delta^3(\mathbf{p}-\mathbf{p}^{\prime})\delta_{ss^{\prime}}\delta_{cc^{\prime}}\\\label{eq11}
\label{11}[a^{c_g}_{\mathbf{k}\lambda},a^{{c^{\prime}_g}^{\dagger}}_{\mathbf{k}^{\prime}\lambda^{\prime}}]&=&(2\pi)^{3}\delta^{3}(\mathbf{k}-\mathbf{k}^{\prime})\delta^{c_g,c^{\prime}_g}\delta_{\lambda\lambda^{\prime}}
\end{eqnarray}

The non-relativistic approximation function for hybrid and mesons are described as follows :
\begin{widetext}
\begin{align}\label{eq12}
\nonumber|A(L_g,L_{q\bar{q}},S_{q\bar{q}},J_A,M_{J_A})\rangle=\sum\int\frac{d^3\mathbf{p}_1d^3\mathbf{p}_2d^3\mathbf{k}}{(2\pi)^9}(2\pi)^3\chi^{\mu}_{ss^\prime}\frac{\lambda^{c_{g}}_{c_q,c_{\bar{q}}}}{4} 
 \psi_{L_{q\bar{q}}M_{L_{q\bar{q}}}}\left(\frac{m_{\bar{q}}\mathbf{p}_1-m_q\mathbf{p}_2}{m_q+m_{\bar{q}}}\right)\;
\psi_{L_{g}M_{L_g}}\left(\frac{\mathbf{k}(m_{\bar{q}}+m_{\bar{q}})-(\mathbf{p}_1+\mathbf{p}_2)\omega_g}{(m_q+m_{\bar{q}})\omega_g}\right)\;\\
\langle L_{g},M_{L_{g}};1,\lambda_{g}|J_{g},M_{J_{g}}\rangle
\langle L_{q\bar{q}},M_{L_{q\bar{q}}};J_{g},
 M_{J_{g}}|L,m^{\prime}\rangle
\langle L,m^{\prime};S_{q\bar{q}},M_{S_{q\bar{q}}}|J_A,M_{J_{A}}\rangle
\delta^{3}(\mathbf{p}_1+\mathbf{p}_2+\mathbf{k}-\mathbf{P}_A)
 b^{\dagger}_{\mathbf{p}_1s_1c_1}d^{\dagger}_{\mathbf{p}_2s_2c_2}a^{{c_g}^{\dagger}}_{\mathbf{k}\lambda_g}|0\rangle.
\end{align}
\end{widetext}
The final $B$ meson's state is given by:
\begin{align}\label{eq13}
\nonumber |B&(L_B,S_B,J_B,M_{J_B})\rangle=\sum_{M_{L_B},M_{S_B}}\int \frac{d^3\mathbf{p}_1d^3\mathbf{p}_3}{(2\pi)^6\sqrt{3}}(2\pi)^3\chi^{\mu}_{ss^\prime}\\
&\langle L_{B},M_{L_{B}};S_{B},
\nonumber M_{S_{B}}|J_{B},M_{J_{B}}\rangle\delta^3(\mathbf{p}_1+\mathbf{p}_3
-\mathbf{P}_{B})\;\\
&\times \psi_{L_B M_{L_{B}}}\left(\frac{m_{\bar{q_{i}}}\mathbf{p}_1-m_q\mathbf{p}_3}{m_q+m_{q_{i}}}\right)\;
b^{\dagger}_{\mathbf{p}_1s_1}d^{\dagger}_{\mathbf{p}_3s_3}|0\rangle.
\end{align}
The spatial wave function form of another final state meson $C$ is consistent with $B$. Here, $\psi_{{L}M_{L}}$ represent the spatial wave functions, which usually is taken as the simple harmonic oscillator wavefunction.

By combining the initial and final state wave functions with the Hamiltonian from Eq. (\ref{eq10}), it is straightforward to obtain the matrix element $\langle
BC|H_I|A\rangle=g\ (2\pi)^3 \ \delta^{3}(\mathbf{P}_A-\mathbf{P}_B-\mathbf{P}_C)\ M_{\ell J}(A\rightarrow BC)$ for the hybrid state $A$ decaying into mesons $B$ and $C$, which has been converted to partial wave amplitudes \cite{Ding:2006ya}. The expression is as follows:
\begin{widetext}
\begin{align}\label{eq14}
\nonumber M_{\ell J}&(A\rightarrow BC)=\sum_{\begin{array}{cccccc}
M_{L_g},&\lambda_g,&M_{L_{q\bar{q}}},&M_{S_{q\bar{q}}},&M_{L_B},&M_{S_{B}},\\
M_{L_C},&M_{S_C},&M_{J_B},&M_{J_C},&M_{J},&M_{\ell}
\end{array}}
{\cal C}\;{\cal F}
{\cal S}(M_{S_{q\bar{q}}},\lambda_g,M_{S_B},M_{S_{C}})
\nonumber  I(M_{L_{q\bar{q}}},M_{L_{g}},M_{L_B},M_{L_{C}},M_{\ell})\\
&\nonumber \langle L_g,M_{L_g};1,\lambda_{g}|J_{g},M_{L_{g}}+\lambda_g\rangle
\langle L_{q\bar{q}},M_{L_{q\bar{q}}};J_{g},M_{L_{g}}+\lambda_g|L,M_{L_{g}}
\nonumber+\lambda_g+M_{L_{q\bar{q}}}\rangle
\langle L,M_{L_{g}}+\lambda_g+M_{L_{q\bar{q}}}; S_{q\bar{q}},
M_{S_{q\bar{q}}}|J_A, M_{J_A}\rangle \\
&\langle L_B,M_{L_B};S_{B}, M_{S_B}|J_B,M_{J_B}\rangle\langle L_C,M_{L_C};S_C,M_{S_C}|J_C,M_{J_C}\rangle\langle
J_B,M_{J_B};J_{C},M_{J_C}|J,M_J\rangle
\langle\ell,M_{\ell};J,M_J|J_A,M_{J_A}\rangle.
\end{align}
\end{widetext}
Here, ${\cal C}=\frac{2}{3}$ is the color overlap factor,  ${\cal F}$ is the flavor overlap factor, $ {\cal S}$ is the spin overlap factors and $I(M_{L_{q\bar{q}}},M_{L_{g}},M_{L_B},M_{L_{C}},M_{\ell})$ is the spatial overlap factor.

The spin overlap factor is:
\begin{align}\label{eq15}
\nonumber{\cal S}&(M_{S_{q\bar{q}}},\lambda_g,M_{S_B},M_{S_C})=\sum_{S}\sqrt{6(2S_B+1)(2S_C+1)(2S_{q\bar{q}}+1)}\\
&\nonumber\times\left
\{\begin{array}{lll}
\frac{1}{2}&\frac{1}{2} & S_B\\
\frac{1}{2}&\frac{1}{2} & S_C\\
S_{q\bar{q}}& 1 & S
\end{array}
\right\}\langle
S_{q\bar{q}},M_{S_{q\bar{q}}};1,
\lambda_g|S,M_{S_{B}}+M_{S_C}\rangle \\
&\langle S_B,M_{S_B};S_C,M_{S_C}|S,M_{S_{B}}+M_{S_C}\rangle.
\end{align}

 Eq. (\ref{eq15}) indicates a decay selection rule, the “spin selection rule”: if the $q\bar{q}$ in hybrid is in a net spin single configuration then decay into final states consistent only of spin single states is forbidden, which also  proposed in
 Ref. \cite{Page:1998gz}.   The flavor overlap factor ${\cal F}$ is:
\begin{eqnarray}\label{eq16}
{\cal F}=\sqrt{(2I_B+1)(2I_C+1)(2I_{A}+1)}\left
\{\begin{array}{lll}
i_1& i_3 & I_B\\
i_2&i_{4} & I_C\\
I_A& 0 & I_A
\end{array}
\right\}\eta \ \varepsilon,
\end{eqnarray}
where I (i) represents the isospin quantum number corresponding to the hadron (quark),
 $\eta=1$ if the gluon goes into strange quarks and
$\eta=\sqrt{2}$ if gluon goes into non-strange ones, which means the ability of gluon to annihilate into non-strange quarks is stronger than that of annihilating into strange quarks.
 $\varepsilon$ is
the number of the diagrams contributing to the decay. Finally the
spatial overlap is given by:
\begin{align}
\nonumber\label{17}I(&M_{L_{q\bar{q}}},M_{L_{g}},M_{L_B},M_{L_{C}},M_{\ell})=\int\frac{d^3\mathbf{p}d^3\mathbf{k}}{\sqrt{2\omega_g}(2\pi)^6}\;\\
&\times\psi_{L_{q\bar{q}} M_{L_{q\bar{q}}}}(\mathbf{P}_B-\mathbf{p})
\psi_{L_g M_{L_g}}(\mathbf{k})\;
\psi^*_{L_B M_{L_B}} \left(\frac{m_{\bar{q_{i}}}\mathbf{P}_B}{m_q+m_{\bar{q_{i}}}} -\mathbf{p}-\frac{\mathbf{k}}{2}\right)\nonumber\\
&\times\psi^{*}_{L_C
M_{L_C}}\left(-\frac{m_{q_{i}}\mathbf{P}_B}{m_{q_{i}}+m_{\bar{q}}}+\mathbf{p}-\frac{\mathbf{k}}{2}\right)\;d\Omega_{B}{Y^{M_{\ell}}_{\ell}}^{*}(\Omega_{B}).
\end{align}
The calculation results indicate that the presence of the spatial integral term leads to the decay width of the excited state being lower than that of ground state.

The partial decay width is \cite{Ding:2006ya}:
\begin{equation}
\label{eq20}\Gamma_{\ell J}(A\rightarrow
BC)=\frac{\alpha_s}{\pi}\frac{p_BE_BE_C}{M_A}|M_{\ell
J}(A\rightarrow BC)|^2.
\end{equation}
where $\alpha_s \approx 0.7\pm0.3$, which represents the infrared quark-gluon vertex coupling \cite{Luna:2006tw,Natale:2006nv,Aguilar:2001zy}.

In our calculations, the parameters are set as follows  \cite{Ding:2006ya}:
$\beta_{q\bar{q}}=0.3\rm{GeV}$, $m_s=0.55\rm{GeV}$, $m_u=m_d=0.33\rm{GeV}$, $\omega_g=0.8\rm{GeV}$, and $\beta_B=\beta_C=\beta_g=0.4\rm{GeV}$. 

Selection rules are pivotal in determining the decay processes of hybrid mesons. For instance, the spin selection rule prohibits decay into final states composed entirely of spin-singlet mesons if the hybrid’s $q\bar{q}$ configuration is in a spin-singlet state. Additionally, spatial overlap integrals, which describe the overlap of wavefunctions between initial and final states, typically lead to suppressed decay widths for excited states relative to ground states.
 
Another decay selection rule indicates that, under the assuming $\beta_B = \beta_C$, the decay of the GE state into $S+S$ is forbidden \cite{Iddir:2000yb}. This rule arises from the treatment of integrals under the condition $\beta_B = \beta_C$. In Ref. \cite{Iddir:1988jd}, the integrals are evaluated with $\beta_B \neq \beta_C$, and the results indicate that the spatial overlap is generally proportional to $(\beta_B - \beta_C)^2$, which is typically small. The same outcome is observed in the flux tube model \cite{Close:1994hc}. Similarly, if the two S-wave states possess different internal structures or sizes, as discussed in Ref. \cite{Close:1994pr}, the selection rule prohibiting $S+S$ as final states no longer applies.


\section{RESULTS AND ANALYSIS }\label{sec4}

\subsection{The $I^GJ^{PC}=1^-1^{-+}$ state}

The combined analysis of masses and decay widths provides essential insights into the nature of exotic mesons, particularly those with $J^{PC}=1^{-+}$. Given the quantum numbers $I^GJ^{PC}=1^-1^{-+}$, we assume that $\pi_1(1600)$ corresponds to the configuration $(u\bar{u}-d\bar{d})g/\sqrt{2}$ in our calculations. For the decay channels involving $f_1(1285)$ and $f_1(1420)$, we model them as follows:
\begin{align}\label{eq23}
     \begin{pmatrix}
    |f_1(1285)\rangle\\
    |f_1(1420)\rangle
    \end{pmatrix}&= \begin{pmatrix}
   \cos\alpha & -\sin\alpha \\ \sin\alpha & \cos\alpha
    \end{pmatrix}
    \begin{pmatrix}
    |n\bar{n}\rangle\\
    |s\bar{s}\rangle
    \end{pmatrix},
\end{align}
where $\alpha \approx 30^\circ$ \cite{Dudek:2013yja}.

Assuming $\pi_1(1600)$ as a hybrid state, its total decay width is calculated to be 65 MeV, with $b_1(1235)\pi$ identified as the primary decay channel (see Table \ref{table2}). However, this prediction exhibits a significant discrepancy with Experimental measurements \cite{ParticleDataGroup:2024cfk}. In the current framework, the $D$-wave contribution to the $b_1(1235)\pi$ decay channel is found to vanish, restricting the decay mechanism exclusively to $S$-wave processes. This result is directly conflicts with the PDG-reported branching ratio $\mathcal{B}((b_1\pi)_{D\text{-wave}})/\mathcal{B}((b_1\pi)_{S\text{-wave}}) = 0.3 \pm 0.1$ \cite{ParticleDataGroup:2024cfk}. The observed inconsistency challenges the hybrid state interpretation and strongly suggests the need to explore alternative configurations, such as tetraquark states and/or molecular formations.

\begin{table}[htbp] 
\caption{Partial and total decay widths of the $\pi_1(1600)$ state (in MeV $\times \alpha_s$). The final states include all charge conjugate pairs.}
\label{table2}
\renewcommand\arraystretch{1.2}
\begin{tabular*}{86mm}{@{\extracolsep{\fill}}cccc} 
\toprule[1pt]\toprule[1pt]
$b_1(1235)\pi$ & $f_1(1285)\pi$ & Total & Experimental \cite{ParticleDataGroup:2024cfk} \\ \cline{1-4}
56.6            & 8.4             & 65   & $370^{+50}_{-60}$ \\                                                   
\bottomrule[1pt]\bottomrule[1pt]
\end{tabular*}
\end{table}

In a parallel investigation, $\pi_1(2015)$, while exhibiting a mass consistent with lattice QCD predictions for the first hybrid excitation, displays decay widths that conflict sharply with experimental constraints, suggesting that it could potentially occupy a distinct class of exotic mesons. Carrying the exotic quantum numbers $I^G J^{PC} = 1^-1^{-+}$, $\pi_1(2015)$ with a PDG reported mass of $2014 \pm 20_{-16}^{+16}$ MeV \cite{ParticleDataGroup:2024cfk} lies within 7$\%$ of the 2193 MeV hybrid state mass predicted in lattice QCD (see Table \ref{table01}). Building upon theoretical foundations established by \cite{Meyer:2010ku} and corroborating LQCD mass spectra \cite{Dudek:2010wm}, the mass agreement provides strong theoretical support for assigning $\pi_1(2015)$ as the  first radial excitation of $1^-1^{-+}$ hybrid. Under this assignment, we postulate the flavor wavefunction $(u\bar{u}-d\bar{d})g/\sqrt{2}$ for $\pi_1(2015)$, enabling systematic computation of its decay properties. For the decay channels involving $K_1(1270)\bar{K}$ and $K_1(1400)\bar{K}$, we treat them as linear combinations of $^1P_1$ and $^3P_1$ states \cite{Page:1998gz,Kokoski:1985is}, 
\begin{equation} \label{eq21}
\begin{array}{l} 
|K_1(1270)\rangle = \sqrt{\frac{2}{3}}|K_1(^1P_1)\rangle + \sqrt{\frac{1} {3}}|K_1(^3P_1)\rangle\\
|K_1(1400)\rangle =  -\sqrt{\frac{1}{3}}|K_1(^1P_1)\rangle +  \sqrt{\frac{2}{3}}|K_1(^3P_1)\rangle
\end{array}
\end{equation}
and the decay channels associated with $h_1(1170)$, $h_1(1380)$, we regard them as 
\begin{align}\label{eq23}
     \begin{pmatrix}
    |h_1(1170)\rangle\\
    |h_1(1380)\rangle
    \end{pmatrix}&= \begin{pmatrix}
   \sin\alpha_i & \cos\alpha_i \\  \cos\alpha_i & -\sin\alpha_i 
    \end{pmatrix}
    \begin{pmatrix}
    |n\bar{n}\rangle\\
    |s\bar{s}\rangle
    \end{pmatrix},
\end{align}
where $\alpha_i=78.7^\circ $ \cite{Wang:2019qyy}.

From our theoretical framework, the total decay width of $\pi_1(2015)$ is determined as 3.21 MeV (Table \ref{table3}), demonstrating a sharp conflict with the PDG reported upper limits $\Gamma >100$ MeV \cite{ParticleDataGroup:2024cfk}. The primary decay channel of $\pi_1(2015)$ is $b_1(1235)\pi$, while other decay channels can be neglected. Experimentally, $\pi_1(2015)$ has been observed in the $b_1(1235)\pi$ channel. However, the narrow decay width leads us to question the interpretation of $\pi_1(2015)$ as the first excited state of the $1^-1^{-+}$ hybrid. To date, $\pi_1(2015)$ production has been exclusively detected in $b_1(1235)\pi$ and $f_1(1285)\pi$ final states \cite{E852:2004rfa,E852:2004gpn}, with no confirmed observations of hadronic transitions to other channels.

\begin{table}[htbp]
\caption{The partial and total widths of the strong decays of the $\pi_1(2015)$ state.(width is in MeV $\times \alpha_s$ for the channels)} \label{table3}
\renewcommand\arraystretch{1.2}
\begin{tabular*}{86mm}{@{\extracolsep{\fill}}ccccc}
\toprule[1pt]\toprule[1pt]
$b_1(1235)\pi$   & $f_1(1285)\pi$     & $h_1(1170)\rho$     & $b_1(1235)\omega$     & $K_1(1270)\bar{K}$    \\
  1.75           & 0.23               & $\approx0$           &  $\approx0$             & $\approx0$         \\
$K_1(1400)\bar{K}$     & $a_1(1260)\rho$     & $\eta(1295)\pi$    & $\eta(1475)\pi$      & $\eta_2(1645)\pi$    \\                $\approx0$        & $\approx0$          & $0.4$               &  0.12                & 0.3               \\ 
$\rho(1450)\pi$  & $\rho(1700)\pi$    & $\pi(1300)\eta$     &                       &                      \\               $0.31 $           & $\approx0$         & $0.1$              &                      &                \\ 
\cline{4-5}
               &                   &                     &                  &               \\
               &                   &                     &   Total          &  Exp.~\cite{ParticleDataGroup:2024cfk}    \\
               &                   &                     &   3.21          &  230$\pm$32$^{+73}_{-73}$     \\                                
\bottomrule[1pt]\bottomrule[1pt]
\end{tabular*}
\end{table}

Furthermore, we predict the properties of the first excited isoscalar hybrids $\pi_1(2193)$ with $I^GJ^{PC}=1^-1^{-+}$.  Its decay properties are listed in Table \ref{table4}. This state is anticipated to exhibit narrow decay width , with dominant decay channels such as $K_1(1650)\bar{K}$ and $b_1(1235)\pi$. For the decay channel involving $K_1(1650)\bar{K}$, we model $K_1(1650)$ as a linear combination of the $2^1P_1$ and $2^3P_1$ states, based on meson mass predictions in Ref. \cite{Godfrey:1985xj}

\begin{table}[htbp]
\caption{The partial and total widths of the strong decays of the $\pi_1(2193)$ which is assumed as the first excited state of $1^-1^{-+}$  hybrid.(width is in MeV $\times \alpha_s$ for the channels)} \label{table4}
\renewcommand\arraystretch{1.2}
\begin{tabular*}{86mm}{@{\extracolsep{\fill}}ccccc}
\toprule[1pt]\toprule[1pt]
$b_1(1235)\pi$   & $f_1(1285)\pi$     & $h_1(1170)\rho$     & $b_1(1235)\omega$     & $K_1(1270)\bar{K}$    \\
  5.4           & 0.81               & $0.1$               &  $\approx0$            & 0.13         \\
$K_1(1400)\bar{K}$     & $a_1(1260)\rho$     & $\eta(1295)\pi$    & $\eta(1475)\pi$      & $\eta_2(1645)\pi$    \\              $\approx0$      &$\approx0$          & $0.86$               &  0.37                & 1.51               \\ 
$\rho(1450)\pi$  & $\rho(1700)\pi$    & $\pi(1300)\eta$     & $K(1460)\bar{K}$           & $\pi(1300)\rho$       \\  
$0.9 $           & 0.25               & $0.4$               & 0.16                &   $0.16$            \\ 
$a_1(1260)\eta$   & $f_1(1420)\pi$    & $K_1(1270)\bar{K}^*$   & $K_1(1650)\bar{K}$         &                \\
  0.2            & 0.12               & $\approx0$           & $20.1$                    &                \\
\cline{4-5}
               &                   &                     &                  &               \\
               &                   &                     &   Total          &  Exp.        \\
               &                   &                     &   31.47          &  $-$     \\            
                                                                 
\bottomrule[1pt]\bottomrule[1pt]
\end{tabular*}
\end{table}


\subsection{The ground states of $ I^GJ^{PC}=0^+1^{-+}$}

As discussed in Section \ref{sec2}, the mass of the pure $s\bar{s}g$ state is 2023 MeV, which exceeds 1855 MeV, while the pure $(u\bar{u} + d\bar{d})g/\sqrt{2}$ state has a mass below 1855 MeV. This indicates that $\eta_1(1855)$ is likely a mixture of $s\bar{s}g$ and $(u\bar{u} + d\bar{d})g/\sqrt{2}$ states, as proposed by \cite{Swanson:2023zlm}. This hypothesis also supports efforts to construct the $1^{-+}$ nonet, as explored in \cite{Chen:2023ukh, Qiu:2022ktc, Eshraim:2020ucw}.

As proposed in Refs. \cite{Chen:2023ukh,Eshraim:2020ucw,Shastry:2023ths}, we assume that $\eta_1(1855)$ is the higher state and $\eta_1(1600)$ is the lower state of the $1^{-+}$ hybrid isoscalar nonet. In this paper, we follow this assumption.
The mixing can be parameterized as follows:
\begin{align}\label{eq23}
     \begin{pmatrix}
    |\eta_1^H\rangle\\
    |\eta_1^L\rangle
    \end{pmatrix}&= \begin{pmatrix}
    \sin\theta & \cos\theta\\ \cos\theta & -\sin\theta
    \end{pmatrix}
    \begin{pmatrix}
    |n\bar{n}g\rangle\\
    |s\bar{s}g\rangle
    \end{pmatrix},
\end{align}
$\theta$ is the mixing angle and $n\bar{n}=(u\bar{u}+d\bar{d})/\sqrt{2}$.

\begin{figure}
    \centering
    \includegraphics[width=1\linewidth]{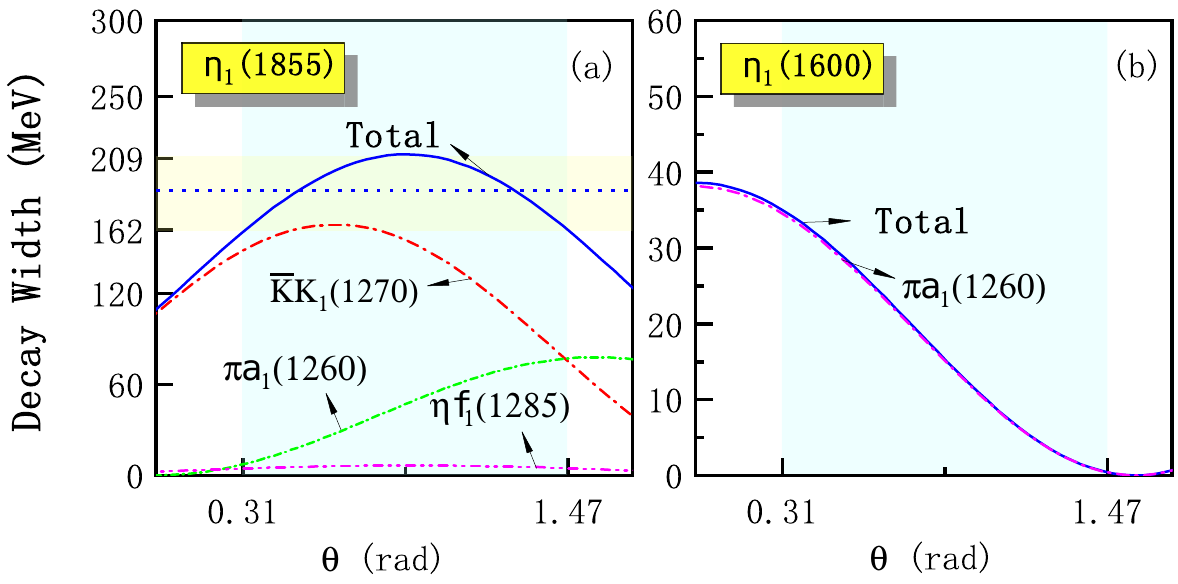}
    \caption{The decay width of $\eta_1(1855)$ and $\eta_1(1600)$ which was assumed as a higher state and lower state, respectively.  (a) represents total and partial decay of $\eta_1(1855)$, (b) represents total and partial decay of $\eta_1(1600)$. The yellow shaded area represents the range of the experimental decay width. The blue shaded area denotes the range of theoretical mixing angles.}
    \label{1855+1600}
\end{figure}

In contrast, if $\eta_1(1855)$ is interpreted as a higher isospin partner, its properties strongly suggest a mixed hybrid state. Table \ref{table1855high} summarizes its decay results. The dominant decay channels, $a_1(1260)\pi$ and $K_1(1270)\bar{K}$, align closely with theoretical predictions (see Fig. \ref{1855+1600}(a)), and the mixing angle between $(u\bar{u}+d\bar{d})g/\sqrt{2}$ and $s\bar{s}g$ is constrained to the range of $17.7^\circ$ to $84.2^\circ$. These findings provide compelling evidence for classifying $\eta_1(1855)$ as a hybrid meson and underscore the significance of mixing effects in its decay dynamics.

For quantum numbers $I^GJ^{PC}=0^+1^{-+}$, the state $(u\bar{u}+d\bar{d})g/\sqrt{2}$ is assigned as the lower isoscalar partner to the $\eta_1(1855)$ state, with the assumption that $\eta_1(1855)$ is the higher isoscalar partner. The lower state is labeled as $\eta_1(1600)$, with the mixing angle ranging from $17.7^\circ$ to $84.2^\circ$, as previously discussed. The results presented in Table \ref{table8} show that $a_1(1260)\pi$ is the characteristic decay channel. The total decay width of $\eta_1(1600)$, as a function of the mixing angle, is shown in Figure \ref{1855+1600} (b).

The results emphasize the necessity for precise experimental measurements of decay branching ratios, especially for $\alpha_s$. Such measurements will further constrain theoretical models and elucidate the hybrid nature of the particle. Additionally, targeted searches for the predicted excited states are essential for advancing our understanding of the $1^{-+}$ hybrid meson spectrum.

\begin{table}[htbp]
\caption{The partial and total decay widths of the $\eta_1(1855)$, which was assumed as a higher state. $s \equiv  \sin\theta, c  \equiv  \cos\theta$. (width is in MeV $\times \alpha_s$ for the channels)} \label{table1855high}
\renewcommand\arraystretch{1.2}
\begin{tabular*}{86mm}{@{\extracolsep{\fill}}ccc}
\toprule[1pt]\toprule[1pt]
 $a_1(1260)\pi$       &     $f_1(1285)\eta$           &  $\pi(1300)\pi$    \\   
 $ 77.9s^2 $          &  $2.5c^2 +4.1s^2+6.4cs$         &  $4.7s^2$   \\ 
                      &                             &   $K_1(1270)\bar{K}$ \\
                      &                             & $106.7c^2 +58.8s^2 +158.4cs$\\
\cline{2-3}
                   &  Total                          &  Exp.~\cite{ParticleDataGroup:2024cfk}     \\
                   & $109.2c^2+145.5s^2+164.8cs$     &   $188\pm18^{+3}_{-8}$    \\   
                                                                 
\bottomrule[1pt]\bottomrule[1pt]
\end{tabular*}
\end{table}

\begin{table}[htbp]
\caption{The partial and total decay widths of the $\eta_1(1600)$, which was assumed as a lower state. $s \equiv  \sin\theta, c  \equiv  \cos\theta$. (width is in MeV $\times \alpha_s$ for the channels) } \label{table8}
\renewcommand\arraystretch{1.2}
\begin{tabular*}{86mm}{@{\extracolsep{\fill}}ccccc}
\toprule[1pt]\toprule[1pt]
 $a_1(1260)\pi$          &  $\pi(1300)\pi$     &     & Total          &  Exp. \\
  $38.1c^2$              &   $0.5c^2$         &     & $38.6c^2$      &  $-$   \\
\bottomrule[1pt]\bottomrule[1pt]
\end{tabular*}
\end{table}

\subsection{The first excited states of $ I^GJ^{PC}=0^+1^{-+}$}

\begin{figure}
    \centering
    \includegraphics[width=1\linewidth]{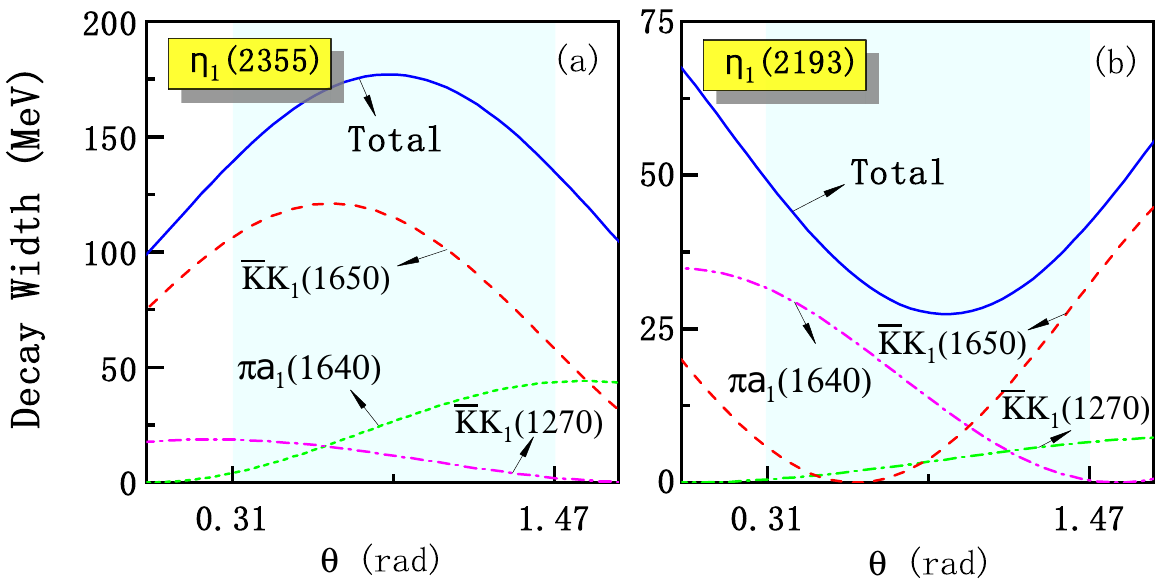}
    \caption{The decay width of $\eta_1(2355)$ and $\eta_1(2193)$ predicted by this work.  (a) represents total and partial decay of $\eta_1(2355)$ , (b) represents total and partial decay of $\eta_1(2193)$. The shaded area denotes the range of theoretical mixing angles.}
    \label{2355+2193}
\end{figure}

In accordance with the mass spectrum calculated in Section \ref{sec2}, the first excitation state of $\eta_1(1855)$ is labeled as $\eta_1(2355)$, maintaining the same mixing angle as that of $\eta_1(1855)$. The decay width results are presented in Table \ref{table7}. For a clearer visualization of the decay widths, refer to Fig. \ref{2355+2193}(a). The total decay width of $\eta_1(2355)$ is approximately 150 MeV, with dominant decay channels including $K_1(1270)\bar{K}$, $K_1(1650)\bar{K}$, and $a_1(1640)\pi$.

Additionally, the first excited state of $\eta_1(1600)$ is named $\eta_1(2193)$, serving as the partner of $\eta_1(2355)$. The decay widths are presented in Table \ref{table9}. As shown in Fig. \ref{2355+2193} (b), the main decay modes of $\eta_1(2193)$ include $K_1(1270)\bar{K}$, $K_1(1650)\bar{K}$, and $a_1(1640)\pi$. We anticipate that future experiments will investigate these channels to detect $\eta_1(2355)$ and $\eta_1(2193)$.

\begin{table}[htbp]
\caption{The partial and total widths of the strong decays of the $\eta_1(2355)$ , which was assumed as a higher state. $s \equiv  \sin\theta, c  \equiv  \cos\theta$. (width is in MeV $\times \alpha_s$ for the channels)}  \label{table7}
\renewcommand\arraystretch{1.2}
\begin{tabular*}{86mm}{@{\extracolsep{\fill}}ccc}
\toprule[1pt]\toprule[1pt]
$\eta(1295)\eta$               & $\eta(1475)\eta$                &  $\pi(1300)\pi$                      \\
 $0.49s^2+0.77c^2+1.23cs$         & $\approx0$                    &  $8.4s^2 $                       \\
 
 $K(1460)\bar{K}$                     & $K^*(1410)\bar{K}$              & $a_1(1260)\pi$                          \\
 $0.49s^2+3.4c^2+2.6cs$         & $0.31s^2+2.1c^2+1.6cs$           &$7.5s^2$                                \\
 
 $f_1(1285)\eta$                 &  $f_1(1420)\eta$            & $a_1(1640)\pi$                            \\
$\approx0$                       & $\approx0$                  &$44s^2$                              \\

 $K_1(1270)\bar{K}$                   &  $K_1(1400)\bar{K}$                & $h_1(1170)\omega$                        \\
 $1s^2+17.7c^2+8.4cs$           &$\approx0$                    &  $0.75s^2$                                  \\

 $K_1(1650)\bar{K}$                     &  $K_2(1770)\bar{K}$             & $b_1(1235)\rho$                             \\
 $46s^2+75c^2+117.4cs$            & $\approx0$                 & $1.5s^2$                                \\

$\eta_2(1645)\eta$                & $\pi_2(1670)\pi$              &                                  \\
 $0.53s^2$                         &  $11s^2$                    &                                  \\
                                  &                               &                                   \\
\cline{2-3}
                           &    Total                            &  Exp.                                \\
                           &    $121.97s^2+98.97c^2+131.23cs$     &  $-$                               \\  
\bottomrule[1pt]\bottomrule[1pt]
\end{tabular*}
\end{table}

\begin{table}[htbp]
\caption{The partial and total widths of the strong decays of the $\eta_1(2193)$  ,which was assumed as a lower state. $s \equiv  \sin\theta, c  \equiv  \cos\theta$. (width is in MeV $\times \alpha_s$ for the channels)} \label{table9}
\renewcommand\arraystretch{1.2}
\begin{tabular*}{86mm}{@{\extracolsep{\fill}}ccc}
\toprule[1pt]\toprule[1pt]
$K(1460)\bar{K}$                       & $K^*(1410)\bar{K}$                   & $\pi(1300)\pi$                    \\
$0.16c^2+1.1s^2-0.84cs$           &  $0.12c^2+0.8s^2-0.62cs $           & $4.75c^2$                \\

 $K_1(1270)\bar{K}$                    & $K_1(1650)\bar{K}$                       &  $K_1(1400)\bar{K}$               \\
 $ 7.5s^2$                       & $20c^2+38s^2-55.2cs$                & $\approx0$                  \\
 
 $\eta(1295)\eta$               & $a_1(1260)\pi$                       & $a_1(1640)\pi$                   \\
$0.2c^2+0.33s^2-0.51cs$          &$3.66c^2 $                           & $34.8c^2$                       \\
 
$\pi_2(1670)\pi$                &                                   &                                 \\
  $3.8c^2 $                     &                                   &                                   \\
 
                                  &                                   &                               \\
\cline{2-3}
                           &    Total                                 &  Exp.                           \\
                           &   $67.49c^2+47.73s^2-57.17cs$           &  $-$                   \\  
\bottomrule[1pt]\bottomrule[1pt]
\end{tabular*}
\end{table}

\section{Discussion and conclusion}\label{sec5}

This study investigates the properties of $J^{PC}=1^{-+}$ hybrid mesons and their isoscalar counterparts through comprehensive mass and decay analyzes. Using a potential model inspired by the lattice QCD and the constituent-gluon framework, we calculate the masses and decay widths of key exotic mesons, shedding light on their structure and classification.

Our findings indicate that $\eta_1(1855)$ is a strong candidate for a mixed hybrid meson, characterized by $(u\bar{u}+d\bar{d})g/\sqrt{2}$ and $s\bar{s}g$ components, with mixing angles constrained to the range of $17.7^\circ$ to $84.2^\circ$. Its mass and decay properties, particularly the dominant channels $a_1(1260)\pi$ and $K_1(1270)\bar{K}$, closely align with theoretical predictions. These results strongly support the classification of $\eta_1(1855)$ as a hybrid meson.

In contrast, the narrow decay widths predicted for $\pi_1(1600)$ and $\pi_1(2015)$ deviate significantly from experimental observations, challenging their classification as hybrid mesons. Alternative explanations, such as tetraquark or molecular configurations, merit further investigation. This study also predicts the excited states of $\eta_1(1855)$ and $\eta_1(1600)$, specifically $\eta_1(12355)$ and $\eta_1(2193)$, respectively. Their primary decay channels are identified as $K_1(1650)\bar{K}$, $K_1(1270)\bar{K}$, and $a_1(1640)\pi$. However, the relatively narrow width of $\eta_1(2193)$ presents a challenge for experimental detection.

This work emphasize the importance of precise experimental measurements to resolve ambiguities in hybrid meson classification. Future experiments should focus on measuring branching ratios for key decay channels, such as those of $\eta_1(1855)$, to refine mixing angle estimates. Furthermore, searches for the predicted excited states will be instrumental in advancing our understanding of hybrid meson dynamics and their role in exotic hadronic states.

The results presented in this study demonstrate the complex relationship between theoretical predictions and experimental investigations in the exploration of exotic mesons. Continued collaboration between experimental efforts and theoretical advancements, including refined lattice QCD calculations and improved decay models, will be critical in uncovering the true nature of hybrid mesons and enhancing our understanding of the strong interaction.

\section*{Acknowledgements}

This work is supported by  the National Natural Science Foundation of China under Grant No.~12405104 and  No.~12305087, the Start-up Funds of Nanjing Normal University under Grant No.~184080H201B20, the Natural Science Foundation of Hebei Province under Grant No.~A2022203026, the Higher Education Science and Technology Program of Hebei Province under Contract No.~BJK2024176, and the Research and Cultivation Project of Yanshan University under Contract No.~2023LGQN010.


\end{document}